\documentclass[12pt]{article}
\usepackage{amsmath}
\usepackage{amssymb}
\begin{document}
\title{Gaussian Effective Potential \\
        in Light Front $\phi^4_{1+1}$}

\author{
G.~B.~Pivovarov\\
{\it Institute for Nuclear Research,}\\
{\it Moscow, 117312 Russia}\\
{\it E-mail: gbpivo@ms2.inr.ac.ru}
}

\date{September 14, 2004}
\maketitle

\abstract{Gaussian effective potential is obtained for $\phi^4_{1+1}$
quantized on a light front. It coincides with the one obtained previously
within the equal time quantization. The computation of the paper substantiates
the claim that light front quantization reproduces the phase structure
of the theory implied by the equal time quantization.}

\newpage

Gaussian effective potential (GEP) can be computed nonperturbatively for any
theory whose Hamiltonian is a polynomial in canonical variables. Its meaning
is discussed in \cite{Stevenson:1985zy}, where it was demonstrated that, 
e.g., for
$\phi^4_{1+1}$, GEP has nontrivial minima at nonzero value of the field that
become absolute minima beyond a critical value of the coupling. The critical
value of the coupling predicted by GEP for $\phi^4_{1+1}$ is in agreement
with the critical value obtained in the lattice computation 
\cite{Loinaz:1997az}. The treatment in \cite{Stevenson:1985zy} 
is performed within
the equal time quantization.

There is an alternative approach to quantization of fields 
possessing a number of advantages. It is the so-called light front 
quantization (for a review, see \cite{Brodsky:1997de}). One of the 
objections against this scheme of quantization is that it has troubles
in reproducing the known facts about phase structure of quantum field 
theories. A good test case is the theory $\phi^4_{1+1}$, because it is
proved rigorously for this theory that there is a spontaneous breaking of the
reflection symmetry $\phi\rightarrow -\phi$. It takes place when
the coupling goes beyond a critical value.

It can be demonstrated heuristically that the phase transition persists
in $\phi^4_{1+1}$ under the light front quantization (see \cite{Kim:2003ha}). 
In
\cite{Kim:2003ha}, Chang's reasoning \cite{Chang:1976ek} was extended to the
light front quantization. This reasoning implies the presence
of a phase transition, but does not predict a critical value of the coupling.

In this note, we continue the line of \cite{Kim:2003ha}, and demonstrate that
GEP obtained under the light front quantization for 
$\phi^4_{1+1}$ coincides with the one obtained under the equal time 
quantization. 

Let us recall the derivation of GEP for $\phi^4_{1+1}$ in equal time 
quantization. The derivation begins with the expression for the 
Hamiltonian density,
\begin{equation}
\label{ham}
{\mathcal H}(x) = \frac{1}{2}\pi^2(x) + 
\frac{m^2}{2}\phi^2(x) + \frac{g}{4}\phi^4(x).
\end{equation}
Here $\pi(x)$ and $\phi(x)$ are the canonical variables.
Next we decompose $\pi$ and $\phi$:
\begin{equation}
\phi(x) = \phi_0 + \int\frac{dk}{\sqrt{4\pi\omega(k)}}\big
[a(k)e^{-ikx} + a^{\dagger}(k)e^{ikx}\big],
\end{equation}
\begin{equation}
\pi(x) = \int\frac{dk}{i\sqrt{4\pi}}\sqrt{\omega(k)}\big
[a(k)e^{-ikx} - a^{\dagger}(k)e^{ikx}\big].
\end{equation}
Here $\phi_0$ is a constant, and $\omega(k)$ is an even function of $k$.
Regardless of the value of $\phi_0$ and behavior of $\omega(k)$, canonical 
commutation relation between $\phi$ and $\pi$ implies the 
canonical commutator $[a(l),a^\dagger(k)]=\delta(l-k)$.

Next step is to compute the expectation of ${\mathcal H}$ with respect 
to the vacuum annihilated by $a(k)$:
\begin{eqnarray}
<{\mathcal H}(x)>& =& \int\frac{dk}{8\pi}
\big(\omega(k) + \frac{m^2}{\omega(k)}\big) + 
3g\Big[\int\frac{dk}{8\pi\omega(k)}\Big]^2 \nonumber\\
&+&3g\phi^2_0\int\frac{dk}{8\pi\omega(k)} 
+ \frac{m^2}{2}\phi^2_0 + 
\frac{g}{4}\phi^4_0.
\end{eqnarray}
We now seek for $\omega(k)$ that would minimize the above vacuum
expectation at fixed $\phi_0$. Requiring variation of the expectation
with respect to $\omega(k)$ to vanish, we obtain the equation for
$\omega(k)$:
\begin{equation}
\label{omega}
\omega^2(k) = m^2 + k^2 + 3g\big(\phi^2_0 + \int\frac{dk}{4\pi\omega(k)}\big).
\end{equation}
GEP is the above expectation of the Hamiltonian density taken at the
solution to Eq. (\ref{omega}). It is a function of $\phi^2_0\equiv R$ 
(the variable $R$ is introduced for later convenience). We denote 
this function $V(R)$. 

To get rid of an (infinite) constant, let us consider the derivative
$\partial V(R)/\partial R$,
\begin{equation}
\label{derivative}
\frac{\partial V(R)}{\partial R} = 
\frac{m^2}{2} + 3g\int\frac{dk}{8\pi\omega(k)} + \frac{g}{2} R
\end{equation}
(to obtain this, one should notice that due to Eq. (\ref{omega}) 
the dependence of $\omega$ on $R$ can be ignored in the derivation of 
the rhs). The value of this derivative at $R=0$ equals by definition half
of the renormalized mass squared:
\begin{equation}
\label{renorm}
m^2_r = m^2 + 3g\int\frac{dk}{4\pi\bar\omega(k)},
\end{equation}
where $\bar\omega$ is $\omega$ at $R=0$.

We now express Eqs. (\ref{omega}) and (\ref{derivative}) in terms of $m_r$
(the aim is to get rid of the ultraviolet divergences):
\begin{eqnarray}
\omega^2(k)& =& \mu^2(R) + k^2,\label{1}\\
\frac{\partial V(R)}{\partial R}& =& \frac{\mu^2(R)}{2} - gR,\label{2}\\
\mu^2(R)&\equiv& m^2_r + 3g\Big(\int\frac{dk}{4\pi}
\Big[\frac{1}{\omega(k)}-\frac{1}{\bar\omega(k)}\Big] + R\Big).\label{3}
\end{eqnarray}
In the last line we introduced a ``mass'' $\mu(R)$. It is a function
of $R$, and coincides with $m_r$ at $R=0$. Performing the integration in
$k$ explicitly in the definition of $\mu(R)$ (we can do it because of the
simple dependence of $\omega$ and $\bar\omega$ on $k$), we obtain the equation
for $\mu(R)$:
\begin{equation}
\label{mass}
\mu^2(R) = 1 - \frac{3g}{4\pi}\log\mu^2(R) + 3gR.
\end{equation} 
Here and from now on we measure all the dimensionfull quantities in 
the units where $m_r=1$. The last equation implies that $\mu^2(R)$ 
is a growing function of positive $R$; its growth starts from the
value $\mu^2(R=0) = 1$.

Finally, we can integrate the derivative of $V$ in $R$ to obtain 
the explicit expression for $V(R)$:
\begin{equation}
\label{gep}
V(R) = -\frac{gR^2}{2} + \frac{\mu^2(R)-1}{6g}
\Big[\frac{\mu^2(R)-1}{2} + 1 + \frac{6g}{8\pi}\Big].
\end{equation}
The last two equations determine $V(R)$ unambiguously in accord with
\cite{Stevenson:1985zy}. For properties of this $V(R)$, see 
\cite{Stevenson:1985zy}.

Let us now repeat the above derivation for the light front quantization.
Specifically, we use the scheme suggested in \cite{Kim:2003ha}. In this paper,
a regularization was introduced in the Lagrangian of the theory, and 
Hamiltonian quantization was applied to the regularized theory with the
initial conditions set at a fixed value of the light front time
$x^+=(x^0+x^1)/\sqrt{2}$. The regularization involves two parameters,
the dimensionless parameter $\epsilon$, and the mass parameter $M$. 
The regularization is removed when $\epsilon$ vanishes and $M$ goes to
infinity. The resulting Hamiltonian density is
\begin{equation}
\label{lfham}
{\mathcal H}_{lf}(x)=\frac{1}{2}p^2(x)+\frac{\epsilon}{2}\phi_-^2(x)
+\frac{m^2}{2}\phi^2(x) + \frac{g}{4}\phi^2(x).
\end{equation}
Here $p(x) \equiv (\pi(x)-\phi(x))/\sqrt{\epsilon-4\partial^2/M^2}$,
and $\phi_-\equiv\partial\phi$. The derivative is in 
$x^-\equiv (x^0-x^1)/\sqrt{2}$.
As before, $\pi$ and $\phi$ are the canonical variables. The variable $x$ 
is now $x^-$ (for the equal time quantization, $x$ was identical to $x^1$).

This Hamiltonian is quite different from the one appearing in equal time
quantization (see Eq. (\ref{ham})). For example, the kinetic term 
involving $p^2(x)$ is formally divergent when the regularization is removed
at fixed canonical variables. We find it to be a remarkable fact that
GEP implied by this Hamiltonian coincides with the standard one
of Eq. (\ref{gep}) after the regularization is removed.

Repeating literally the above derivation of GEP starting from Eq. (\ref{lfham})
for the Hamiltonian, one observes that the only modification of 
Eqs. (\ref{1})--(\ref{3}) implied by switching over from the 
Hamiltonian (\ref{ham})
to the Hamiltonian (\ref{lfham}) takes place in the equation 
expressing $\omega$ in terms of momentum and mass. For the light front 
Hamiltonian the expression is as follows:
\begin{equation}
\label{lfomega}
\omega^2_{lf}(k) = 
k^2 + (\epsilon k^2 + \mu^2(R))(\epsilon + \frac{4k^2}{M^2}).
\end{equation}

At first glance, the difference in the dependence of $\omega_{lf}$ on $k$
with respect to the one taking place for the equal time quantization
may cause a difference in the equation for $\mu(R)$, and, consequently,
may change the equal time expression for GEP. But this is not the case. To see
it, switch to the variable $k_{lf} = k/\sqrt\epsilon$ in the integrals over
$k$ involved in Eq. (\ref{3}), and neglect all the terms formally disappearing
in the limit $\epsilon\rightarrow 0$, $M\rightarrow\infty$. After this, 
the integral in $k_{lf}$ involved in Eq. (\ref{3}) becomes identical to 
the one appearing in equal time quantization. Therefore, the final
expression for GEP (Eq. (\ref{gep})) is reproduced in light front quantization.

We conclude with the following observations. First, under the light front
quantization GEP is formed by the modes of low momentum. Specifically, 
the characteristic scale of the momenta in the integrals 
$\int\,dk/\omega_{lf}(k)$ is of the order $\sqrt{\epsilon}$, and, in the
limit of the regularization removed, the characteristic momenta vanish.
Second, negative momentum modes are as important for forming GEP as the
modes with positive momentum. As discussed in \cite{Kim:2003ha}, the modes 
with negative momenta correspond to tachyons under the light front 
quantization. Thus, one cannot ignore tachyons in the computation
of GEP. This is in contrast to perturbative computations, where all the
integrations in momenta can be restricted to positive momenta, and,
therefore, tachyons can be ignored.

This work was supported in part by RFBR grant no. 03-02-17047.

\end{document}